\documentclass[12pt]{iopart}
\usepackage{iopams}
\usepackage{graphicx}
\input amssym.def \input amssym

\begin{document}

\title[]{Twelve-dimensional Pauli group contextuality }

\author{Michel Planat$^{1}$
}

\address{ $^1$ Institut FEMTO-ST, CNRS, 32 Avenue de l'Observatoire, F-25044 Besan\c con, France.}
\ead{michel.planat@femto-st.fr}



\begin{abstract}

The goal of the paper is to check whether the real eigenstates of the observables in the single qudit Pauli group may lead to quantum contextuality, the property that mutually compatible and independent experiments depend on each other. We find that quantum contextuality crops up at dimension twelve in various configurations with a few rays. We use the Shannon capacity for characterizing the corresponding orthogonality graphs.
  Some arithmetical properties underlying the qudit contextuality are outlined.

\end{abstract} 


\section{Introduction}

Quantum contextuality is the statement that the result of an experiment depends on the other mutually compatible measurements that may potentially be performed. This surprising non-classical (and non-local) property was first pointed out by Kochen and Specker (the KS-theorem) from a judiciously choosen set of $117$ rays/vectors of the real three-dimensional space. Later, it was simplified to a highly symmetric arrangement of $33$ rays attached to cube of edge size $\sqrt{2}$ \cite[fig. 7.2, p. 198]{Peres}. In this pionnering work, a context means a maximal set of mutually compatible/commuting observables. In essence, a KS-proof is the impossibility to assign the value $1$ ($1$ means true) to one single ray in a complete orthonormal basis, while preserving the identity of the ray, and that of the other rays, in all separate (complete) bases. In particular, no two orthogonal rays can both be assigned $1$.

The less stringent option that the rays do not necessary belong to a complete orthonormal basis, but to maximal sets of mutually orthogonal rays, was emphasized in a recent paper \cite{YU2011}. In the following, a proof of quantum contextuality consists of a finite sets of rays that cannot be each assigned the values $0$ or $1$ in such a way that (i) in any group of mutually orthogonal rays, not all the rays are assigned the value $0$ and (ii) no two orthogonal rays are both assigned the value $1$ \cite{Aravind1998}.

Yu and Oh proof of state-independent contextuality (SIC) only needs $13$ rays attached to a cube of edge size $1$ as follows
\begin{equation}
V=\{100,010,001,011,01\bar{1},101,10\bar{1},110,1\bar{1}0,\bar{1}11,1\bar{1}1,11\bar{1},111\}.
\end{equation}
Let us index the rays consecutively from  $1$ to $13$, i.e. $1\equiv 100$, $2\equiv 010\ldots$ The following orthogonality relations between triples and pairs of vectors are satisfied
\begin{eqnarray}
& (1,2,3),\nonumber \\
& (1,4,5),~~~~(4,11),(4,12),(5,10),(5,13),\nonumber \\
& (2,6,7),~~~~(6,10),(6,12),(7,11),(7,13),\nonumber \\
& (3,8,9),~~~~(8,10),(8,11),(9,12),(9,13). \nonumber\\\nonumber
\end{eqnarray}
It is straigthforward to check that, while it is possible to assign the value $1$ to one and only one ray of the four complete bases (triads) above, while preserving the identity of the other rays that are henceforth assigned the value $0$, it is not possible to do so with all the remaining orthogonal  pairs. The end result of an assignment $1$ to a single ray in all four triads immediately leads to a contradiction to non-contextuality because two orthogonal rays in a pair may both be assigned $1$. In addition, an experimentally testable inequality involving the $13$ projectors associated to the rays is found to be satisfied by all non-contextual hidden variable models while being violated by all three-level states \cite[eq. (4)]{YU2011}.

Yu and Oh's orthogonality graph is easily shown to be non-planar, of automorphism group $\mathbb{Z}_2^2 \rtimes S_3 $ (the semidirect product of the Klein group by the three-letter symmetric group), and of chromatic number $\kappa=4$. In a recent paper \cite{Cabello2011}, it is proposed that a SIC proof with an arrangement of  $N$ rays in dimension $q$ is such that the orthogonality graph of the rays has chromatic number $q+1$, and conversely, see \cite{Cabello2011,Cabello2011bis} for the ins and outs of the latter statement.

To finalize this introduction, let us remind that there exist even smaller proofs of quantum contextuality that are state-dependent, as the $8$-ray set \cite[fig. 2]{Clifton1993} (a subset of the Yu and Oh's set $V$)
\begin{equation}
W=\{010,001,101,10\bar{1},110,1\bar{1}0,\bar{1}11,111\}.
\end{equation}
The orthogonality relations may be easily infered from the $13$-ray set above as
\begin{equation}
(2,3),(2,6,7),(6,10),(7,11),(7,13),(3,8,9),(8,10),(9,13).
\label{8vertex}
\end{equation}
The corresponding graph is planar, of automorphism group $\mathbb{Z}_2^2$ and its faces consist of two triangles, two pentagons and one heptagon. The Shannon capacity of the $8$-ray set is found to be that of the heptagon (see Sec. \ref{heptagon} for the definition). The reason why the pentagon and the $8$-vertex configuration (\ref{8vertex}) are basic building blocks of the Kochen-Specker theorem and related quantum paradoxes is given in \cite{CabelloPRL,Badziag2011} (see also \cite{Waegell2011}).

In the present paper, inspired by the recent progress about quantum contextuality with a small number of rays in the three-dimensional space \cite{Peres}-\cite{Clifton1993}, and by the proofs of the Kochen-Specker theorem in the four-dimensional space \cite{Arends2011}-\cite{Aravind2011}, we asked ourselves what is the smallest configuration of a single qudit system, with rays belonging to the real eigenstates of the corresponding Pauli group, leading to quantum contextuality? A qudit may be seen as $q$-level system, i.e. a qubit (two levels) corresponds to spin $\frac{1}{2}$, a qutrit (three levels) corresponds to spin $1$, a quartit (four levels) corresponds to a spin $\frac{3}{2}$ system, an octit (eight levels) correponds to a spin $\frac{7}{2}$ system and so on; these spin systems may be manipulated experimentally by applying selective radio frequency pulses at the resonance frequency between two energy levels.

In the present paper, we find that one needs at least $12$ levels for displaying quantum contextuality of a single qudit. In particular, we single out and characterize contextual sets of $11$ and $9$ rays that form non-planar orthogonality graphs with non trivial Shannon capacity \footnote{This research was made possible following our recent efforts to understand the arithmetical and geometrical structure of low dimensional qudits \cite{Planat2011} and our extensive use of the Magma software \cite{Bosma1997}.}.

\section{Qudit twelve-dimensional quantum contextuality}




A single $q$-dimensional qudit is defined by a Weyl pair $(X,Z)$ of {\it shift} and {\it clock} cyclic operators satisfying 
\begin{equation}
ZX-\omega XZ =0,
\label{Weylpair}
\end{equation}
where $\omega=\exp \frac {2i\pi}{q}$ is a primitive $q$-th root of unity and $0$ is the null $q$-dimensional matrix. In the standard computational basis $\{\left|s\right\rangle, s \in {\mathbb{Z}_q}\}$, the explicit form of the pair is as follows
\begin{equation}
X=\left(\begin{array}{ccccc} 0 &0 &\ldots &0& 1 \\1 & 0  &\ldots & 0&0 \\. & . & \ldots &.& . \\. & . & \ldots &.& . \\0& 0 &\ldots &1 & 0\\ \end{array}\right),~~ Z= \mbox{diag}(1,\omega,\omega^2,\ldots,\omega^{q-1}).
\label{Paulis}
\end{equation}

The Weyl pair generates the single qudit Pauli group $\mathcal{P}_q=\left\langle X,Z\right\rangle$, of order $q^3$, where each element may be written in a unique way as $\omega^aX^bZ^c$, with $a,b,c \in \mathbb{Z}_q$.

The study of commutation relations in a arbitrary single qudit system may be based on the study of symplectic modules over the modular ring $\mathbb{Z}_q^2$. Let us define a {\it isotropic line} as a set of $q$ points on the lattice $\mathbb{Z}_q^2$ such that the symplectic product of any two of them is $0 (\mbox{mod}~ q)$. To such an isotropic line corresponds a maximal commuting set in the quotient group $\mathcal{P}_q/Z(\mathcal{P}_q)$, where $Z(\mathcal{P}_q)$ is the center of the group \cite{Planat2011}-\cite{ShalabyVourdas2012}. There exist $\sigma(q)$ isotropic lines/maximal commuting sets \cite[eq. (5)]{Planat2011}-\cite[eq. (3)]{PlanatAnselmi}, where $\sigma(q)$ is the sum of divisor function. One finds a total amount of  $N_q=q \sigma(q)$ distinct vectors/rays as eigenstates of the operators on the isotropic lines, a formula still to be established rigorously \footnote{This is the sequence A064987 of Sloane's encyclopedia of integer sequences.}.

In the following we are interested in the real subset of the aforementioned rays and the orthogonal graph they form. It is not a trivial task to get it, it is why we made an extensive use of the Magma software \cite{Bosma1997} for the matrix calculations, the extraction of the maximum cliques of the commutation graph between the operators within the Pauli group, the search of the eigenstates of the operators, the filtering of the real eigenstates and the derivation of their orthogonality graph. Then, we derived a program similar to the one created in \cite[prog. state01.C in sect. 2.2]{Pavicic2005} for detecting a possible quantum contextuality of the orthogonality graph.

We analyzed all real vectors of qudit systems of dimension $q<24$. Only dimension $12$ turns out to provide a case of quantum contextuality that we now exhibit. In the 12-dimensional qudit system, there are $q \sigma(q)=336$ vectors arising from the $\sigma(q)=28$ maximum cliques (of equal size $q=11$) of the Pauli graph. Among all vectors, $48$ are real. The real vectors decompose according to their degree in the orthogonality graph as
$$8+4+24+12,$$
where $12$ vectors have degree $38$, $24$ have degree $34$, $4$ vectors have degree $28$ and the remaining $8$ have degree $20$. According to our calculations, quantum contextuality is present in this ray system and we try to reduce it as much as possible so as to get a simple analytical proof. Removing the $8+12$ vectors of degrees $20$ and $38$, one is left with $28$ vectors that form a total amount of $37$ maximal cliques of orthogonal vectors, $8$ of them of size $12$ and the remaining ones of smaller sizes $4$ to $8$. Note that the decomposition $28=24+4$ mimicks the decomposition of the number of isotropic lines of operators $\sigma(12)=28$ into $\psi(12)=24 $ operators that form the projective line $\mathbb{P}_1(\mathbb{Z}_{12})$ and $4$ outliers \cite[table 1]{Planat2011}. Here $\psi(q)$ is the Dedekind psi function, see \cite[table 1]{Planat2011}. Quantum contextuality is observed in the orthogonality graph of the $28$ rays and the chromatic number is $12$. Finally, one proceeds with a step by step reduction of the $18$ rays to $11$ rays still preserving the contextuality. Our final building block is as follows
\begin{equation}
\{e_3,e_4,e_5,e_7,e_8,e_{12},(010\bar{1})^3,(10\bar{1}0)^3,(001)^4,(100)^4,0100\bar{1}00100\bar{1}0\},
\label{ray11}
\end{equation}
where $e_i$ is a vector of the computational basis that is, the coordinate is $1$ at position $i$ and $0$ elsewhere, and the exponent in the expression of the ray means repetition of the coordinates inside the brackets.
In the following, the rays are numbered consecutively from $1$ to $11$. The orthogonality relations between the rays are given from the maximum cliques
\begin{eqnarray}
&( 1, 2, 3, 4, 5, 6 ),~( 1, 3, 4, 7 ),~( 1, 3, 5, 6, 10),~( 1, 3, 7, 10),\nonumber \\
&( 1, 2, 4, 6, 11),~( 1, 4, 7, 11),~( 1, 7, 10, 11),~( 1, 6, 10, 11), \nonumber \\
&( 7, 8, 9, 10, 11 ),~( 5, 8, 9, 10),~( 2, 5, 8, 9 ),~( 2, 8, 9, 11 ),\nonumber \\
&( 2, 5, 6, 8 ),~(2, 6, 8, 11 ),~(5, 6, 8, 10),~( 6, 8, 10, 11),\nonumber \\
&( 2, 3, 4, 5, 9 ),~( 2, 4, 9, 11),~( 3, 4, 7, 9 ),\nonumber \\
&( 4, 7, 9, 11),~( 3, 7, 9, 10 ),~( 3, 5, 9, 10 ).\nonumber \\ \nonumber
\label{rays11bis}
\end{eqnarray}
The proof of contextuality is straigthforward. Let us assign the value $1$ (true) to the first ray $1$ in the first basis. Then, the remaining rays have to be assigned $0$, and similarly for the rays distinct from the ray $1$ in the following seven bases. As a result, in the very last base, the ray $9$ has to be assigned $1$ (true), the other rays in the $10$ bases containing ray $9$  have to be assigned $0$ and it becomes impossible to assign $1$ to any of the rays in the  $4$ bases $( 2, 5, 6, 8 ),\ldots~,( 6, 8, 10, 11)$, located at the fourth line of our list, in violation of the condition (i) (of the introduction) required for non-contextuality. The proof proceeds similarly if one chooses to assign $1$ to any of the five rays in the first base.
The proof is minimal in the sense that no further ray can be removed from the eleven ones while preserving the contradiction to non-contextuality.

The orthogonality graph of our $11$-vertex arrangement, now denoted $G_{11}$, contains the disjoint complete graphs $K_6$ (formed by the rays $1$ to $6$) and $K_5$ (formed by the rays $7$ to $11$) and is thus non-planar of chromatic number $6$. The degree sequence decomposes as 11=2+9: rays $7$ and $8$ have degrees $7$ and the remaining rays have degree $8$. The automorphism group is the semidirect product $\mathbb{Z}_2 \rtimes S_3$. The eigenvalues of the graph are the roots of the following characteristic polynomial

$$(x+1)^2(x^2+2x-1)^2(x^2+2x-2)(x^3-8x^2+10)=0.$$
leading to the spectrum 

$$\left[(-1)^2,(\sqrt{2}-1)^2),(-1-\sqrt{2})^2,(\sqrt{3}-1),(-1-\sqrt{3}),1.213,7.847,-1.051\right],$$
in which the last three eigenvalues are approximations of the real roots of the cubic polynomial that factors in the characteristic polynomial.

One may ask if there exists an obvious property of the graph $G_{11}$ that illustrates the contextuality. Although one cannot further reduce the number of rays, one can reduce the number of orthogonality relations necessary for a proof of the contextuality. For this peculiar graph, the contextuality proof relates to Kuratowski's theorem about non-planar graphs, that a finite graph is planar if and only if it does not contain a subdivision of the complete graph $K_5$ or the complete bipartite graph $K_{3,3}$. We already know that $G_{11}$ contains $K_5$ and is indeed non-planar. Let us display a Kuratowski's obstruction $G_{11}^{(K)}$ og $G_{11}$ in the form of the following subset of the orthogonality relations of $G_{11}$ \cite{Bosma1997}
\begin{eqnarray}
&( 1, 2, 3, 5 ),~( 2, 3, 5, 6 ),\nonumber \\
&(1,11),~(11,10),~(10,9),~(9,8),~(8,7),~(7,4),~(4,6).\nonumber \\ \nonumber
\label{rays11ter}
\end{eqnarray}
The non-planar graph $G_{11}^{(K)}$ is found to have the same automorphism group than $G_{11}$ and the chromatic number $4$. Next we denote $\triangle$ the triangle $(2,3,5)$ and the orthogonality relation between pairs of rays as $(a,b)\equiv a-b$. The orthogonality relations among the rays attached to $G_{11}^{(K)}$ take the schematic $9$-gon form
\begin{equation}
\triangle-1-11-10-9-8-7-4-6-\triangle,
\label{tria}
\end{equation}
which indeed, as for the pentagon encountered before, cannot have its vertices assigned $1$ and $0$ such that no edge is assigned $1-1$. If $\triangle$ happened to be a single ray, then (\ref{tria}) would be a planar graph: a true $9$-gon, of chromatic number $3$. 

\subsection*{The Shannon capacity of the graph $G_{11}^{(K)}$}
\label{heptagon}

To further explore the similarity between $G_{11}^{(K)}$ to the $9$-gon graph, we make use of graph theoretical tools first introduced in classical communication theory \footnote{Details about the definitions and theorems can be found in standard textbooks on graph theory or in the appropriate sections of Wikipedia. See also \cite{Bekius2011}}.

We denote $G\square H$, $G \times H$ and $G \boxtimes H=(G\square H)\cup(G \times H)$ the cartesian product, the tensor product and the strong product of graphs $G$ and $H$, respectively. In the following, we are interested in the product graph $G^k=G \boxtimes G \boxtimes \cdots \boxtimes G $ (with $k$ terms in the product) and the size $\alpha(G^k)$ of a maximum independent set (also called the independence number) of $G^k$.

The Shannon capacity $\Theta(G)$ of the graph $G^k$ is the maximum number of $k$-letter messages that can be sent through a channel without a risk of confusion. It is defined by the expression
\begin{equation}
\Theta(G)=\mbox{sup}_k\sqrt[k]{\alpha(G^k)}.
\label{Shan}
\end{equation}
Even for very simple graphs, the Shannon capacity cannot easily be obtained. Important ingredients are the size $\omega(G)$ of the largest clique (also called the clique number) of $G$, the chromatic number $\kappa(G)$ and indeed the independence number $\alpha(G)$. In a perfect graph, the clique number equals the chromatic number of every induced subgraph so that $\kappa(G)=\omega(G)$; this happens if and only if the graph complement $\bar G$  of $G$ is perfect. As a result, it can be shown that $\Theta(G)=\alpha(G)$ for a perfect graph.   

For a general graph, one has the bounds
\begin{equation}
\alpha(G)\le \Theta(G) \le \theta(G),
\label{Shannon}
\end{equation}
where the upper bound $\theta(G)$ is the Lov\'asz number \cite{Lovasz79}. The Lov\'asz number satisfies the \lq\lq sandwich theorem" $\omega(G)\le\theta(\bar{G})\le \kappa(G)$. In the following, we call a graph $G$ {\it non trivial} whenever $\Theta(G)>\alpha(G)$. Specifically, for a $n$-gon $C_n$ (with $n$ odd), one has
$$\theta(C_n)=\frac{n \cos(\pi/n)}{1+\cos(\pi/n)},$$
that is $\theta(C_3)=1$ (for the triangle), $\theta(C_5)=\sqrt{5}$ (for the pentagon), $\theta(C_7)\sim 3.317$ (for the heptagon) and $\theta(C_9)\sim 4.360$ (for the $9$-gon).

For the triangle and the pentagon, the upper bound in (\ref{Shannon}) is tight. For the pentagon, $\Theta(C_5)\ge \mbox{sup}(2,\sqrt{5})\equiv \theta(C_5)$. For higher order $n$-gons, the direct calculation of the Shannon capacity is out of reach. Only the independence numbers of $C_n$ and $C_n^2$ of small $n$-gons can easily be computed. For the heptagon, one gets $\alpha(C_7)=3$, $\alpha(C_7^2)=10$ and from (\ref{Shan}), $\Theta(C_7)\ge \mbox{sup}(3,\sqrt{10})\sim 3.162$. For the $9$-gon, $\Theta(C_9)\ge \mbox{sup}(4,\sqrt{18})\sim 4.242$.

The non-planar graph $G_{11}^{(K)}$ in (\ref{tria}) can be considered close to the $9$-gon since we numerically get
\begin{equation}
\Theta(G_{11}^{(K)})\ge \mbox{sup}(4,\sqrt{18})\sim 4.242.
\end{equation}
At least for the $11$-ray set (\ref{ray11}), considerations about quantum contextuality meet considerations about the Shannon capacity of polygon graphs.

\subsection*{The Shannon capacity of the graph $G_{9}^{(K)}$}
\label{pentagons}

Apart from $11$-ray systems of the type studied at the previous section, there exist $9$-ray contextual sets such as
\begin{equation}
\{e_3,e_7,e_8,(10)^6,(010)^4,(010\bar{1})^3,(001)^2(00\bar{1})^2,(0100\bar{1}0)^2,100\bar{1}00\bar{1}00100\},
\label{ray9}
\end{equation}
with notations similar to the ones used in (\ref{ray11}). The rays are numbered from $1$ to $9$ and are found to obey the orthogonality (and contextual \footnote{The proof proceeds similarly to that of the contextuality of the $11$-ray system (\ref{ray11}).}) relations of a graph denoted $G_9$
\begin{eqnarray}
&(1,2,5,6,8),~(1,5,8,9),~(1,2,3),~(1,3,9),\nonumber \\
&(2,5,7,8),~(2,3,7),\nonumber \\
&(3,4,7,9),~(5,7,8,9),~(4,6).\nonumber \\ \nonumber
\label{rays9bis}
\end{eqnarray}
The spectrum is
$$[0,-1,-2,5.648,1.522,0.288,0.152,-2.440,-2.171.]$$
The orthogonality graph $G_9$ is build upon two disjoint complete graphs $K_6$ (with the rays $1,2,5,6,8$) and $K_4$ (with the rays $3,4,7,9$). It is non-planar, of automorphism group $\mathbb{Z}_2^2$ and of chromatic number $\kappa=5$. 

As before for the $11$-ray system, one can display a Kuratowski's obstruction $G_9^{(K)}$ which preserves the contextuality. It is the following subset of the orthogonality relations of $G_9$
\begin{eqnarray}
&(1,9),~(9,4),~(4,3),~(3,2),~(2,1),\nonumber \\
&(1,9),~(9,8),~(8,7),~(7,5),~(5,1),\nonumber \\
&(5,6),~(6,4),~(4,9),(~9,1),~(5,1),\nonumber \\
&(2,7).\nonumber \\ \nonumber
\label{rays9ter}
\end{eqnarray}
The graph $G_9^{(K)}$ has the chromatic number $3$ and the automorphism group $\mathbb{Z}_2^2$. It contains a planar configuration of three mutually intersecting pentagons, and the pair $(2,7)$ that ensures the non-planarity. 
To label it, one may use an approximation of the (unknown) Shannon capacity as
\begin{equation}
\Theta(G_{9}^{(K)})\ge \mbox{sup}(4,\sqrt{17})\sim 4.123.
\end{equation}
Contrarily to $G_{11}^{(K)}$, the Shannon signature of $G_9^{(K)}$ is not the one of a recognizable planar graph.

Further contextual $9$-ray sets with non-planar and non trivial Shannon capacity exist but are not investigated here. 

\section{Discussion}

Real eigenstates of operators arising from the single qudit Pauli group have been investigated with the viewpoint of quantum contextuality, the constraint that the measurement outcomes of independent experiments have to be compatible.
While contextuality is a common feature of multiple qubit systems, although not related to entanglement, the mere existence of contextuality in a single qudit may appear at first sight surprising because, contrarily to the case of a multiple qudit system, the eigenstates do not need to be shared by more than one operator.

If the dimension $q$ does not contain a square, the contextuality is dismissed because the number of bases is $\sigma(q)=q+1$, each one containing $q-1$ operators, so that all the $q^2-1$ operators in the Pauli group appear once in the bases due to the equality $(q-1)\sigma(q)=q^2-1$. If the dimension contains a square (this starts at the quartit, with $q=2^2$), the number of bases is $\sigma(q)> q+1$ so that $(q-1)\sigma(q)>q^2-1$ , leaving the room to a degeneracy and a possible contradiction to non-contextuality.

As explained in this paper, we found the first instance of quantum contextuality in the single qudit Pauli group at dimension $12=2^23$, in arrangements of $11$ and $9$ real rays not identified before. Further instances occuring at higher dimensions (containing a square) or with complex rays have still to be discovered.

In addition, whether a non trivial Shannon capacity of the orthogonality graph of a ray system (or of one of its subsets) is a useful signature of contextuality is a work hypothesis worthwhile to be explored in future papers.

\end{document}